\documentclass[epj]{svjour}
\usepackage{graphics}
\usepackage{amsmath}
\usepackage{amsfonts}
\usepackage{amssymb}
\usepackage{color}

\newcommand{\qq}{\qquad}
\newcommand{\q}{\quad}

\newcommand{\primo}{^{\scriptscriptstyle\prime}}
\newcommand{\primoprimo}{^{\scriptscriptstyle\prime\prime}}

\newcommand{\ao}{\bar{a}}
\newcommand{\rhoo}{\rho_{0}}
\newcommand{\azero}{a_{0}}

\newcommand{\tzero}{t_{0}}
\newcommand{\rs}{r_S^\odot}

\newcommand{\Hzero}{H_{0}}
\newcommand{\Ss}{\mathcal{S}}

\newcommand{\pc}{\textrm{ pc}}
\newcommand{\s}{H_I}
\newcommand{\sbar}{\bar{H}_I}
\newcommand{\squareB}{\square}
\newcommand{\qo}{q_{0}}

\newcommand{\reff}[1]{(\ref{#1})}
\newcommand{\figref}[1]{Fig. \ref{#1}}
\newcommand{\refeq}[1]{Eq.(\ref{#1})}

\newcommand{\lesssims}{\,{\scriptstyle\lesssim}\,}

\newcommand{\rts}{r_{TS}}

\begin{document}
\title{Cosmological implications of a viable non-analytical f(R) model}
\author{S. Capozziello\inst{1} \and
N. Carlevaro\inst{2} \and
M. De Laurentis\inst{1} \and
M. Lattanzi\inst{3} \and
G. Montani\inst{4,2}
}                     % Do not remove
\institute{
Department of Physics, University of Naples ``Federico II''; INFN - Section of Naples (Italy). \and
Department of Physics, ``Sapienza'' University of Rome (Italy). \and
Department of Physics and Earth Science, University of Ferrara; INFN - Section of Ferrara (Italy). \and
ENEA - C.R. Frascati (Rome), UTFUS-MAG (Italy).}

\date{}
\abstract{Power-law corrections (having the exponent strictly between 2 and 3) to the Einstein-Hilbert action yield an extended theory of gravity which is consistent with Solar-System tests and properly reproduces the main phases of the Universe thermal history. We find two distinct constraints for the characteristic length scale of the model: a lower bound from the Solar-System test and an upper bound by requiring the existence of the matter-dominated era. We also show how the extended framework can accommodate the existence of an early de Sitter phase. Within the allowed range of characteristic length scales, the relation between the expansion rate and the energy scale of inflation is modified, yielding a value of the rate several orders of magnitude smaller than in the standard picture. The observational implication of this fact is that a tiny value of the tensor-to-scalar ratio is expected in the extended framework. The suppression of primordial tensor modes also implies that the inflationary scale can be made arbitrarily close to the Planck one according to the current limits. Finally, an analysis of the propagation of gravitational waves on a Robertson-Walker background is addressed.
 % end of PACS codes
} %end of abstract
\maketitle

\section{Basic statements}\label{uno}

The most immediate generalization of the Einstein-Hilbert (EH) action deals with a function of the Ricci scalar $R$ analytical in the point $R=0$, thus its Taylor expansion holds \cite{cap,cap1}. This approach is equivalent to deal with a polynomial form of $f(R)$ \cite{large,large1,stelle}, whose free parameters are available to fit the observed phenomena on different sectors of investigation. Despite the appealing profile of such a choice, it is worth noting that it is not the most general case, since real (non-integer) exponents of $R$ are, in principle, on the same footing as the simplest case \cite{nonrat}.

In this paper\footnote{
The signature is set as $[\,-,+,+,+\,]$; Greek indices run form $0$ to $3$; Latin indices run from 1 to 3; the Riemann tensor is defined by $R^{\mu}_{\nu\rho\sigma}= \Gamma^{\mu}_{\nu\rho,\,\sigma}-\Gamma^{\mu}_{\nu\sigma,\,\rho}+ \Gamma^{\lambda}_{\nu\rho}\Gamma^{\mu}_{\sigma\lambda}-\Gamma^{\lambda}_{\nu\sigma}\Gamma^{\mu}_{\rho\lambda}$, and the Ricci tensor by \cite{BO83} $R_{\mu\nu}=R^{\lambda}_{\mu\lambda\nu}$; $(...)\primo$ is the derivative with respect to $R$; $\nabla_\mu$ or $(...)_{;}$ is the covariant derivative; the dot $(...)\dot{}$ is the time derivative; $(...)_{,}$ indicates ordinary differentiation; $\square\equiv\nabla^{\mu}\nabla_{\mu}$; we use natural units $c=\hbar=1$ and we define $\chi \equiv 8\pi G$, $G$ being the Newton constant; the subscript $(...)_{\scriptscriptstyle 0}$ denote cosmological quantities measured today.},
we will focus on such an open issue and we will develop a modified theory of the following form: $f(R)=R+qR^n$, where $q$ and $n$ are two free parameters to be constrained at the observational level \cite{tak,tak1,tak2,tak3,tak4,zac}. In \cite{LM09}, it is shown how the model requires rational non-integer values for $n$ and its restriction in the appropriate interval for the physical interpretation at low curvature, \emph{i.e.} $2<n<3$. The explicit solution of the system derived in \cite{LM09} is here re-analyzed confirming the existence of a lower bound for the characteristic length scale defined by means of $q$.

The cosmological implementation of the resulting modified Friedmann dynamics is also addressed as the central theme of this paper. The fundamental equation for the scale factor are investigated for a perfect fluid matter source and the main phases of the Universe evolution, \emph{i.e.} the radiation- and matter-dominated era, are properly reproduced. The main result of the cosmological analysis is to show the existence of an upper bound of $\sim 70\pc$ (weakly dependent on $n$) for the characteristic length scale, which guarantees the correct matter-dominated Universe evolution. Combining this cosmological bound with the Solar-System constraints, we define the allowed values for $q$, thus providing a restricted region of the parameter space in which the model can be implemented both in the weak-field limit, and in the cosmological framework.

A study of the existence of the standard (exponential) inflationary behavior, in our generalized framework, is also provided. We show how an exponential early evolution of the Universe is still present; however, the existence of a minimum length scale implies that the expansion rate is much smaller than the standard one by several orders of magnitude. We then outline how the production of tensor perturbations is much less efficient than usual. This allows to have the inflationary scale arbitrarily close to the Planck one without spoiling the current limits on the tensor-to-scalar ratio, which is expressed as a function of the model parameters and acquires very tiny values.

Finally, we focus on the study of the gravitational wave (GW) propagation on a Robertson-Walker (RW) background. The GW equation is derived, using the standard conformal formalism, in the transverse-traceless gauge. We obtain a general propagation equation for GWs in the case of a flat RW model. Assuming a power-law behavior of the scale factor, we outline how the existence of a maximum length scale implies that the $f(R)$ corrective term can be always neglected during the Universe evolution and thus it does not yield any observational consequence.

Therefore, we get a precise range of the characteristic length scale of the model in which the modified Lagrangian here addressed provides a viable formulation of the gravitational theory without a significant new physics apart from the very small tensor-to-scalar ration in the primordial spectrum. For other examples of correspondingly viable $f(R)$ model, also able to account for some features of the present Universe (such as dark matter or dark energy), see \cite{NO11,CDL11,NO06}.

\section{Non-analytical power-law f(R) model}\label{due}
We consider the following modified gravitational action in the so-called Jordan frame,
\begin{equation}\label{nonan}
\Ss=-\tfrac{1}{2\chi}\;\textstyle{\int}d^{4}x\,\sqrt{-g}\;f(R)\;,\qq
f(R)=R+qR^n\;,
\end{equation}
where $n$ is a non-integer dimensionless parameter and $q<0$ has dimensions $[L]^{2n-2}$, thus we can define the characteristic length scale of the model as $L_q(n)\equiv|q|^{1/(2n-2)}$. If we decompose the corresponding metric in the form $g_{\mu\nu}=\eta_{\mu\nu}+ h_{\mu\nu}$, where $h_{\mu\nu}$ is a small, static perturbation of the Minkowskian metric $\eta_{\mu\nu}$, the vacuum Einstein equations read as in \cite{LM09} and their structure leads us to focus on the restricted region of the parameter space $2<n<3$. In the present model, terms behaving like $\sim h^{2-p}$ (with $p>0$) appear in the dynamical equations and this corrections are taken into account to describe the post-Newtonian behavior obtaining $p=1$ for $n\to2$, while the modified terms become no longer negligible at all.

In the following, we re-adapt the scheme of \cite{LM09} to gain reliable constraints on the viability of the theory. From the analysis of the weak-field limit in the Jordan frame, it can be shown the possibility to find a post-Newtonian solution by solving the Einstein equations up to the next-to-leading order in $h$, \emph{i.e.}, up to $\mathcal{O}(h^{n-1})$ and neglecting the $\mathcal{O}(h^2)$ contributions since $2<n<3$. In particular, taking the most general spherically-symmetric line element in the weak-field limit as $ds^2=-(1+\Phi)dt^2+(1-\Psi)dr^2-r^{2}d\Omega^2$ (where $\Phi$ and $\Psi$ denote the two generalized gravitational potentials), the modified Einstein equations admit the following solutions:
\begin{subequations}\label{pns}
\begin{align}
\Phi\equiv&\Phi_N+\Phi_{M}=-r_S/r+\Phi_n\;r^{2\;\tfrac{n-1}{n-2}}\;,\qq\q\label{pns2}\\
\Phi_n&=R_n(n-2)^{2}\,/\,6(3n-4)(n-1)\;,\\
R_n&=\big[-q\;6n(3n-4)(n-1)\,/\,(n-2)^2\big]^{1/(2-n)}\;,\\
\Psi\equiv&\Psi_N+\Psi_{M}=-r_S/r+\Psi_n\;r^{2\;\tfrac{n-1}{n-2}}\;,\label{pns3}\\
\Psi_n&=R_n(n-2)\,/\,3(3n-4)\;,
\end{align}
\end{subequations}
where we have split $\Phi$ and $\Psi$ into two terms, the Newtonian part $(...)_N$ and a modification $(...)_M$, and $r_S = 2GM$ is the Schwarzschild radius of a central object of mass $M$. 

\begin{figure}
\centering
\resizebox{0.6\hsize}{!}{\includegraphics*{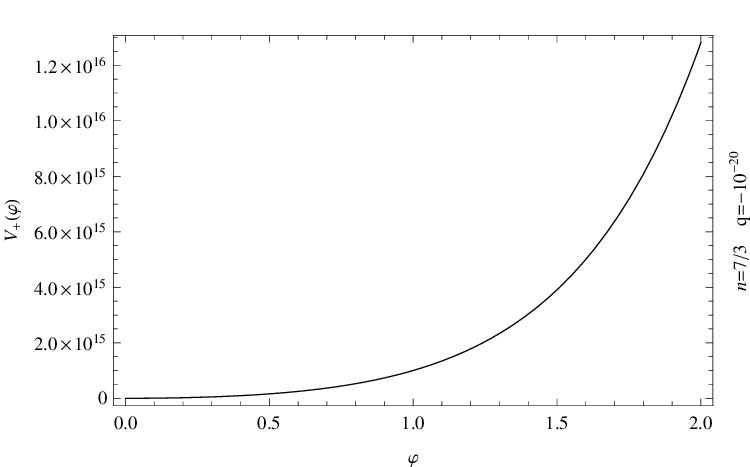}}
\caption{Behavior of $V_+$ described by Eq.\reff{potab} for $n=7/3$ and $q=-10^{-20}$.}
\label{FIGV}
\end{figure}

The gravitational Lagrangian in the Jordan frame can be cast \cite{BS} into an equivalent action for a scalar field $\varphi$ in GR (Einstein frame), by means of a suitable conformal transformation of the metric tensor, \emph{i.e.} $g_{\mu\nu}\to e^{\varphi}g_{\mu\nu}$. This action describes a scalar field $\varphi=-\ln f'(R)$ minimally coupled to the rescaled metric. Although the solution \reff{pns} is well defined for arbitrary values of $2<n<3$, the inflationary paradigm requires (see below) $n=(2\ell+1)/(2m+1)$ (here and in the following $\ell$ and $m$ denotes positive integers). In this case, the potential $V(\varphi)$, describing the scalar field dynamics, reads \cite{LM09}
\begin{align}
\label{potab}
V_{\pm}(\varphi)=\pm\, q\,(1-n)\,e^{2\varphi}\big[(e^{-\varphi}-1)/qn\big]^{n/(n-1)}\;,
\end{align}
which is defined only for $\varphi>0$ \cite{BO83}. The presence of a minimum in the potential is crucial, since the cosmological implementation suggests that it becomes an attractive stable configuration\footnote{A detailed discussion of this point can be found in \cite{tsujikawa} where the Solar-System and equivalence-principle constraints on $f(R)$ gravity are discussed in view of the so called Chameleon Approach \cite{weltman}.}. Indeed, it is easy to show that $V_{-}$ is a monotonically decreasing function of $\varphi$, while $V_{+}$ possesses a minimum in $\varphi=0$. In fact, such a solution starts from this minimum ($V_{+}(0)=0$) and monotonically grows, as depicted in \figref{FIGV}.
%\begin{figure}[!ht]
%\center
%\includegraphics{FIGV}
%\caption{Behavior of $V_+$ described by Eq.\reff{potab} for $n=7/3$ and $q=-10^{-20}$.}
%\label{FIGV}
%\end{figure}

The most suitable arena to evaluate the validity range of the weak-field solutions is the Solar System \cite{zac,tak,tak1,tak2,tak3,tak4}. In this respect, we remark that, as shown in \cite{LM09}, this solution stands also in the presence of matter, allowing the cross-match with the vacuum case. Neglecting the lower-order effects concerning the eccentricity of the planetary orbit, the Solar System test deals with an orbital period $T=2\pi(r/a_c)^{1/2}$ (where $a_c=(d\Phi/dr)/2$ is the centripetal acceleration). Comparing now the correction to the Keplerian period $T_K=2\pi r^{3/2}(GM_\odot)^{-1/2}$, with the experimental data $T_{exp}$ and its uncertainty $\delta T_{exp}$ \cite{BM06}, the correction arising from the $\Phi_M$ term must be imposed smaller than the experimental uncertainty $\delta T_{exp}/T_{exp}\geqslant|T_K-T|/T_K$ and we obtain a lower bound $L_q>L^{Min}_{q}$ for the characteristic length scale \cite{ProcBarca}, as function of $n$. It writes 
\begin{equation}\label{LMin}
L^{Min}_{q}(n)=\Big[\frac{T_{exp}}{\delta T_{exp}}\;\frac{|\Phi_n|}{\rs}\;\frac{n-1}{n-2}\;
r_P^{\tfrac{3n-4}{n-2}}\Big]^{\tfrac{n-2}{2n-2}}\;,
\end{equation}
where $r_P$ is the mean orbital distance of a planet from the Sun and $r_S^\odot$ the Sun Schwarzschild radius.

Evaluating $L^{Min}_{q}$ for all the Solar-System planets using the data of \cite{BM06}, it results that the lower bound for this length scale is maximized with the Pluto parameters, \emph{i.e.} $r_P\simeq 1.91\times10^{-4}\pc$, $T_{exp} \simeq 90\,465 \textrm{ days}$ and $\delta T_{exp}\simeq 10\textrm{ days}$. For $n$ not so close
%\footnote{These values of $n$ can be regarded as typical, in the sense that they are not affected by the peculiarities of the two boundary values.} 
to the limiting cases $n=2$ or $3$, \emph{e.g.} $n\simeq2.66$, one gets $L^{Min}_{q}\sim7.8\times10^{-3} \pc$. The lower bound for $L_q$ does not represent a shortcoming of the model, as we are going to discuss in next Section.

\section{Cosmological constraint for $L_q$}\label{tre}

By assuming a generic time ($t$) power-law behavior of the Universe scale factor $a$, it can be shown how the two main evolutionary (radiation and matter dominated) phases are asymptotically properly reproduced. In particular, toward the singularity (taken in $t=0$), the radiation-dominated scale factor expansion $\sim t^{1/2}$ is recovered for all values of the spatial curvature and a phase of power-law inflation ($\sim t^{n/2}$) can be generated for vanishing curvature. On the other hand, for large $t$, the matter-dominated behavior $\sim t^{2/3}$ is reproduced, neglecting the spatial curvature, as the only asymptotic power-law solution. For a detailed analysis of this modified thermal history of the Universe, see \cite{ProcBarca}.

In a matter-dominated regime where $\Hzero=2/3\tzero$ ($H$ is the Hubble constant), the condition $t\ll382\,\,\tzero$ guarantees to neglect the curvature in the Ricci scalar. Such constraint has been estimated by extrapolating backward in time the standard cosmological parameters. While, the matter solution $a\propto t^{2/3}$ can be asymptotically obtained for $t$ larger than a critical value \cite{ProcBarca}:
\begin{align}
t\gg\;t_c &\equiv\;\mu(n,\qo) \Hzero^{-1}\;,\label{l_tmin}\\
\mu(n,\qo) &\equiv\big|\;\qo\big[-(4/3)^n+2^{(2n+1)}3^{-n}\,n(2n-7/3)\big]\;\big|^{1/2(n-1)}\;,\nonumber
\end{align}
where we have introduced the dimensionless parameter $\qo=\Hzero^{2n-2}q$. In order to preserve the existence of a matter-dominated era lasting as long as in the standard picture, we require that $t_c$ is well before the time of matter-radiation equality $t_{eq}$. To get a numerical estimate, we set $t_{eq}\simeq4 \times10^{-6} \Hzero^{-1}$. Conservatively imposing that $t_c<0.001 t_{eq}$, we obtain $\mu(n,\qo)\leqslant4\times10^{-8}$ (independently of $\Hzero$). This implies an upper limit for $|\qo|$:
\begin{align}
&|\qo|<|\qo|^{Max}(n)=\big[4 \times10^{-8}\big]^{2(n-1)}\big|-(4/3)^n + 2^{(2n+1)}3^{-n}\,n(2n-7/3)\big|^{-1}\;.
\label{q0maxFORMULA}
\end{align}
It is easy to check that the function $|\qo|^{Max}(n)$ is decreasing as $n$ goes from $2$ to $3$ and, in particular, one gets $10^{-31}\lesssims|\qo|^{Max}\lesssims\;10^{-16}$.

\begin{figure}[!ht]
\centering
\resizebox{0.6\hsize}{!}{\includegraphics*{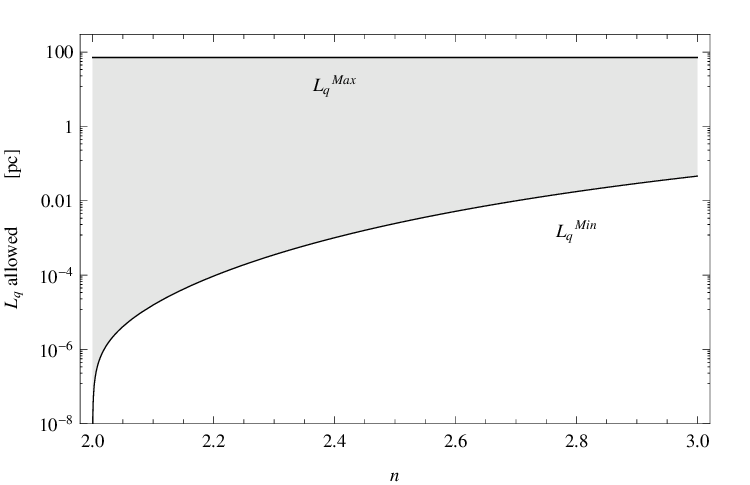}}
\caption{Plot of $L_{q}^{Min}$ (maximized for the Pluto data) of \refeq{LMin} and $L_q^{Max}$, as indicated in the figure.}
\label{length}
\end{figure}

We can thus define the upper cosmological limit for $L_q$ as $L_q^{Max}(n)=[|\qo|^{Max}]^{1/(2n-2)}/\Hzero$ which, considering \refeq{q0maxFORMULA} and using $\Hzero^{-1}\simeq4.2\times10^{9} \pc$, yields to the constraint $52 \pc \lesssim L_q^{Max}\lesssim63 \pc$, for $2<n<3$. The two bounds $L^{Min}_{q}$ (maximized for the Pluto data) and $L_q^{Max}$ are plotted in \figref{length} as a function of $n$.

Summarizing, our analysis states a precise range of validity for the power-law $f(R)$ model we are considering. Indeed, for a typical value of $n$ the fundamental length of the model is constrained to range from the super-Solar-System scale up to a sub-galactic one.

\section{The inflationary paradigm}\label{quattro}

We now address the core analysis of the present paper by characterizing the morphology of the modified inflationary scenario. The most important issue of our study concerns the much smaller value that the e-folding takes with respect to the standard case, when the same vacuum energy density is considered. This key feature implies an high suppression of the tensor-to-scalar ratio generated during the slow-rolling phase. We can thus argue that, in such a modified cosmological Universe, inflation could have happened even very close to the Planckian era.

\subsection{Expansion rate}

We model the inflationary expansion as driven by some component (\emph{e.g.} a scalar field) with negative pressure $p$. We are interested in studying how the action \reff{nonan} affects the de Sitter dynamics and, indeed, if such a phase can exist in the extended framework. Thus, we consider $a=a_I\;e^{\s(t-t_I)}\equiv\ao\;e^{\s t}$, where $\s>0$ is the Hubble constant during the de Sitter phase and $t_I$ denotes the end of inflation. We focus on the case of vanishing spatial curvature and the 00-component of the modified Einstein equations reads as (here $w\equiv p/\rho=const.$ and $\rho=\rhoo\,[a/\azero]^{-3(1+w)}$ is the energy density)
\begin{align}\label{poiajiopajiop}
\ao^{4}\;e^{4\s t}\big[q(-12)^n\;\s^{2n}(1&-n/2)-6\s^{2}\big]+2\tilde{\chi}(\ao\,e^{\s t})^{1-3w}=0\;.
\end{align}
Assuming $w=-1$, \emph{i.e.} $\rho=\rho_I=const.$, Eq.\reff{poiajiopajiop} writes
\begin{align}\label{eq:inflation}
\big[(-1)^{n}12^n\;q(1-n/2)\big]\s^{2n}-6\s^{2}+2\chi\rho_I=0\;.
\end{align}
Here, $\rho_I$ can be assumed of the order of the Grand Unification energy-scale $\rho_I\simeq(10^{16}\,\textrm{GeV})^{4}$ and the standard GR relation $\s=\sbar=\sqrt{\chi\rho_I/3}$ is recovered for $q=0$.

To integrate \refeq{eq:inflation}, we focus on a typical value of the power-law exponent, \emph{e.g.} $n=29/13\sim2.23$. Using \refeq{LMin}, we obtain $L^{Min}_{q}\simeq1.5\times10^{-4}\pc$ and, in turn, $|q|>|q|^{Min}\simeq3.6\times10^{-10}\textrm{pc}^{32/13}$. Let us now fix the parameter $q$ to a reasonable value like $q^{*}\sim-10^{3}|q|^{Min}$ (such assumption will be motivated below). In this case, the solution of \refeq{eq:inflation} is $\s\simeq5.2\times10^{21}\,\textrm{pc}^{-1}\simeq 3.3\times 10^{-11}\,\mathrm{GeV}$. It is worth noting that, for such typical values of $q$, the equation above admits solutions only for $n=(2\ell+1)/(2m+1)$. This constraint is actually the only one emerging in the cosmological implementation of the proposed extended theory.

\begin{figure}[!ht]
\centering
\resizebox{0.42\hsize}{!}{\includegraphics*{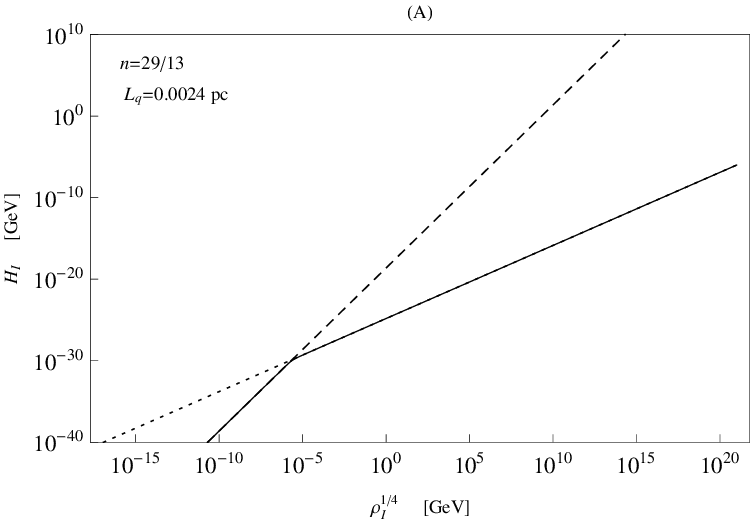}}
\resizebox{0.43\hsize}{!}{\includegraphics*{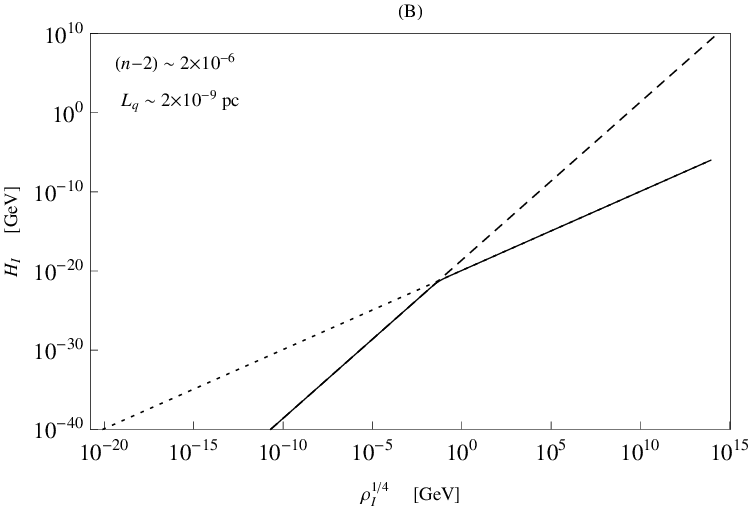}}
\caption{\textbf{(A)}: Numerical solution of Eq.\reff{eq:inflation} with $n=29/13$ and $L_q=2.4\times10^{-3}\,\mathrm{pc}$, corresponding to $q=-10^3|q|^{Min}$ (solid line). We also show the standard solution $\sbar =\sqrt{\chi \rho_I/3}$ (dashed line) and the solution obtained assuming that the $f(R)$ correction term is the leading one (dotted line). \textbf{(B)} The same as the panel A, but for $(n-2)\simeq2\times 10^{-6}$ and $L_q\simeq 2	\times 10^{-9} \mathrm {pc}$ (again corresponding to $q=-10^3|q|^{Min}$).}
\label{fig:Hrho}
\end{figure}

This study shows how an exponential early expansion of the Universe can still be associated to a constant vacuum energy. However, we see that the rate of expansion is significantly lower than the standard one by more than 20 orders of magnitude since $\sbar=\sqrt{\chi\rho_I/3}\simeq3.7\times10^{45}\mathrm{pc}^{-1}\simeq2.4\times10^{13}\,\mathrm{GeV}$. This issue is essentially independent on the adopted value for $\rho_I$ as shown in \figref{fig:Hrho}.
Requiring that the minimum number of e-folds solving the standard paradoxes is the usual one, this implies that the duration of the de Sitter phase should be much longer (time-wise).

\subsection{Spectrum of primordial perturbations}

Let us consider the scalar and tensor fluctuations generated during inflation. Scalar (curvature) perturbations are the seeds from which the density fluctuations, that will grow to become the structures we observe today, are generated. Tensor perturbations $h_{ij}$ are also produced during inflation and correspond to GW fluctuations, giving origin to a cosmological background that could, in principle, leave an imprint in the CMB temperature and polarization spectra. The amplitude of both scalar and tensor perturbations is expressed through their power spectra $P_{\mathcal R}(k)$ and $P_{T}(k)$, quantifying the variance of fluctuations with wave number $k$. In GR, these write
\begin{eqnarray}
P_{\mathcal R}(k)=\big(H^2/2\pi\dot\phi\big)^2\big|_{\,k=aH}\;, \\
P_{T}(k)=8\chi\big(H/2\pi\big)^2\big|_{\,k=aH}\;,
\label{eq:gwpk_st}
\end{eqnarray}
where $\phi$ is the inflaton  and the $k$ dependence emerges because the right-hand side is evaluated at the time the mode of interest leaves the horizon ($k=a H$). Since $H$ is nearly constant during inflation, its value at horizon crossing is, in a first approximation, the same for all $k$-modes; this implies that the spectra should be nearly scale-invariant. We will neglect the scale dependence taking $H=H_I$, and concentrate on the spectral amplitudes.

The form of the perturbation spectra in the framework of modified theories of gravity has been derived in \cite{Hwang:1991mb,Hwang:1996xh,Hwang:2001qk,Hwang:2005hb,capozGW} (see also Sec. 7 of \cite{DeFelice:2010aj} for a summary of the relevant results). Using these results, it can be shown that, in the case under consideration here, the amplitude of the scalar spectrum is unchanged with respect to standard GR, while the tensor amplitude is altered as follows:
\begin{align}\label{eq:prim_spec_GW}
&P_{T}=\big(8\chi/|f'(R)|\big)\;\big[\s/2\pi\big]^2\;,\\
&f'(R)=1+ n q R^{n-1}=1+n\big[4\chi\rho_IL_q^2/(n-2)\big]^{\frac{n-1}{n}},
\end{align}
where we have used $R=12 H_I^2$.

The primordial amplitude of scalar perturbations is well measured by CMB experiments to be $P_{\mathcal R}|_{k=k_0}\simeq 2.2\times 10^{-9}$ at the pivot wave number $k_0=0.05\,\mathrm{Mpc}^{-1}$ (see \cite{Ade:2013zuv,Ade:2013uln}).
Since, as shown above, $H_I$ is suppressed with respect to its standard value, we need to assume that the inflationary potential is very flat, so that $\dot\phi$ is small enough to compensate and give the observed amplitude of scalar fluctuations. 
Very flat, and in fact also exactly flat (at least at tree level) potentials can be achieved for example in the framework of hybrid inflation \cite{Linde:1993cn} or natural inflation \cite{Freese:1990rb} models. In the first case, the potential is flat because the field is trapped in a false vacuum, while in the second case this is due to the shift symmetry of the potential itself. Another possibility to get the right scalar amplitude is simply that $n$ is very close to 2, so that the standard inflationary behavior is recovered.

Now let us turn to the tensor modes. The second term of Eq.\reff{eq:prim_spec_GW} is always much larger than unity; thus we have
\begin{align}
f'\simeq n\Big[\frac{4\chi\,\rho_I\,L_q^2}{n-2}\Big]^{\frac{n-1}{n}}
\!\!\!\!\!=n\Big[\frac{32\pi\,M^4}{(n-2)m_q^2 m_\mathrm{pl}^2}\Big]^{\frac{n-1}{n}}
\!\!\!\gg1\;,
\label{eq:fprime}
\end{align}
where $m_q\equiv L_q^{-1}$ is the energy scale associated to $q$, $M=\rho_I^{1/4}$ the inflation energy scale and $m_\mathrm{pl}\equiv 1/\sqrt{G}\simeq 10^{19}\,\mathrm{GeV}$ the Planck mass. Eq.\reff{eq:fprime} implies that the amplitude of the tensor primordial spectrum is strongly suppressed, first because of the large factor $f'(R)$ in \refeq{eq:prim_spec_GW}, secondly because $H_I$ is itself suppressed with respect to its standard value. Substituting $f'$ and $\s$ into \refeq{eq:prim_spec_GW}, we obtain
\begin{align}
P_{T}%&=\frac{\chi}{12 n \pi^2 \,L_q^2}\left[\frac{n-2}{4\chi\rho_I L_q^2}\right]^{\frac{n-2}{n}}=\nonumber\\
=\frac{4}{3n\pi}\;\frac{m_q^2}{m_\mathrm{pl}^2}\;
\Big[\frac{(n-2)m_\mathrm{pl}^2m_q^2}{32\pi M^4}\Big]^{\frac{n-2}{n}}\;.
\label{eq:PT}
\end{align}
It is interesting to note that, contrarily to the usual case, the amplitude of tensor modes is smaller for large values of $M$, and that in the limit $n\to 2$, it does not depend on $M$ but only on $m_q$, \emph{i.e.} $P_{T}|_{n\to2}\simeq (1/6\pi) (m_q^2/m_\mathrm{pl}^2)$. We now write the tensor spectrum in terms of the prediction of standard GR, given by \refeq{eq:gwpk_st} evaluated for $\s=\sbar$, times a factor $\beta = \beta(n,\,m_q,\,M)$, that will turn out to be very small. Thus, we obtain
\begin{equation}
P_{T} =\frac{(n-2)}{n}^\frac{n-2}{n}\Big[\frac{m_\mathrm{pl}^2 m_q^2}{32\pi M^4}\Big]^\frac{2(n-1)}{n}
\bar{P}_T\equiv\beta\bar{P}_T\;.
\end{equation}
The amplitude of $P_T$ is clearly maximized in the case $m_q=1/L_q^{Min}$, \emph{i.e.} when the Solar-System bounds are saturated. This maximum formally diverges for $n\to 2$ although, for example, for $n=2.01$ we get the tiny value $1/L_q^{Min}=1.4\times10^{-26}\,\mathrm{GeV}$, and it is easy to check that $\beta$ is of order unity when $M$ lies below the MeV scale, that is cosmologically unrealistic. We can conclude that reasonable values of $n$ and $M$ yield $P_T\ll \bar{P}_T$.
\begin{figure}[!ht]
\centering
\resizebox{0.6\hsize}{!}{\includegraphics*{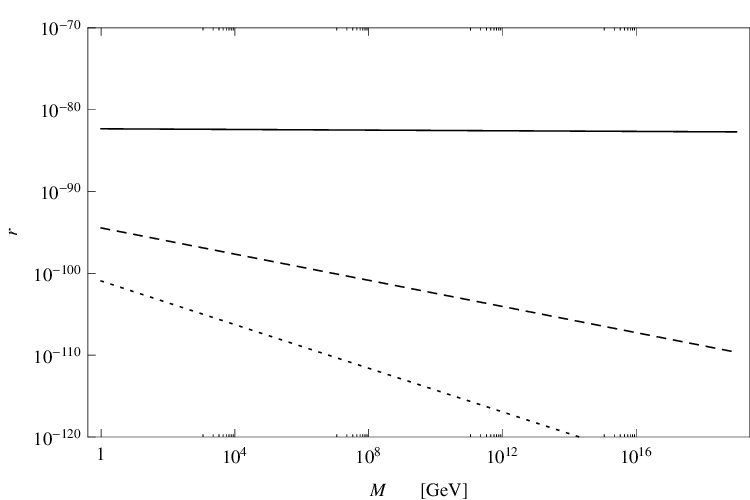}}
\caption{Tensor-to-scalar ratio $\rts$ as function of $M$ (here $m_q$ is assumed to saturate the Solar-System bounds): solid line $n=2.01$; dashed line $n=2.5$; dotted line $n=3$.}
\label{fig:rvsM}
\end{figure}

The amplitude of tensor modes is measured by the ratio $\rts$ between tensor and scalar perturbations, evaluated at some pivot wavenumber $k_0$, \emph{i.e.},  $\rts\equiv (P_T/P_S)|_{k=k_0}$.  The tensor-to-scalar ratio is constrained by the recent Planck data to be $\rts<0.11$ at $k_0 =0.002\,\mathrm{Mpc}^{-1}$ at 95\% confidence level (c.l.) \cite{Ade:2013zuv,Ade:2013uln}.

\begin{figure}[!ht]
\centering
\resizebox{0.4\hsize}{!}{\includegraphics*{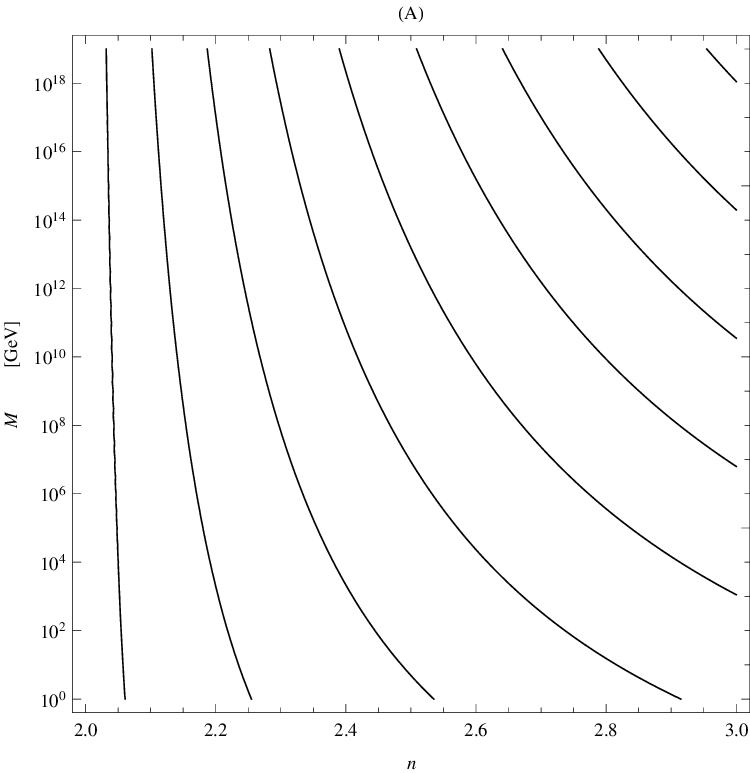}}
\resizebox{0.4\hsize}{!}{\includegraphics*{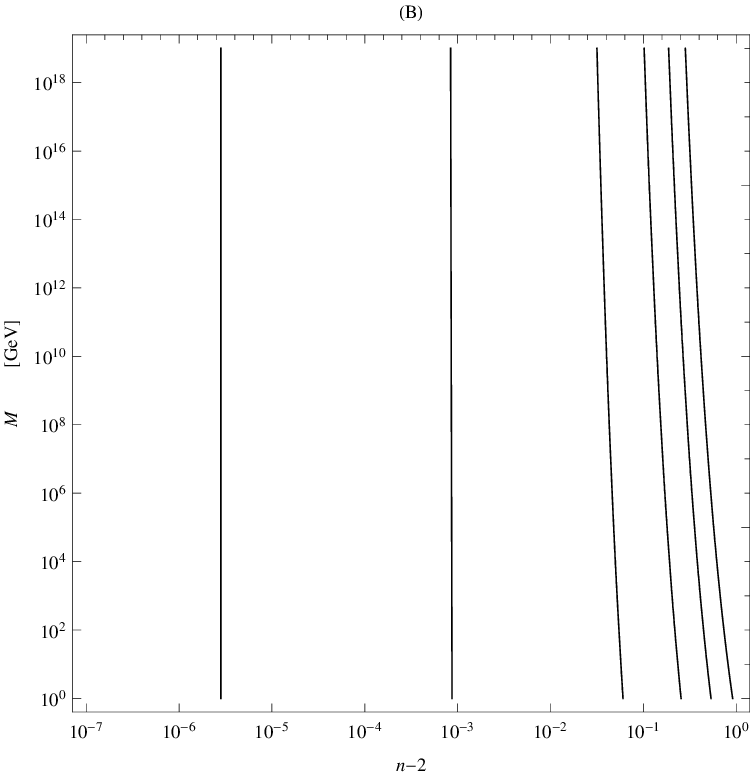}}
\caption{\textbf{(A)} Curves of constant $\rts$ in the $(M,\,n)$ plane. From left to right: $r=\{10^{-85},\, 10^{-90},\,\dots,\,10^{-125}\}$ ($m_q$ is assumed to saturate the Solar-System bounds). \textbf{(B)} The same as panel A in the $(M,\,\log(n-2))$ plane. From left to right: $r=\{10^{-75},\, 10^{-80},\,\dots,\,10^{-100}\}$.}
\label{fig:rvsMn+}
\end{figure}

In our scheme, we obtain $\rts\sim 10^{-83}$ for $n=2.01$ (practically independent of $M$), and still lower values for larger $n$ (see \figref{fig:rvsM}). This implies that the model predicts a GW cosmological background that will be beyond experimental reach even in the far future, at least for the very large wavelengths probed by CMB experiments. In fact, in \figref{fig:rvsMn+} it is evident that potentially detectable values of $\rts$ are allowed in a tiny region of the parameter space with unnaturally small values of $n-2$. More precisely, it is easy to check that, in order to have $\rts\sim10^{-4}$ or larger, $n-2$ has to be of the order of $10^{-41}$ or smaller. The smallness of the tensor-to-scalar ratio can be considered as a precise and falsifiable prediction of the model.

\section{Propagation of gravitational waves}\label{sei}

GW detectors will be of fundamental importance to probe the GR or every alternative theory of gravitation \cite{felix,felix1,felix2,felix3}. A possible target of these experiments is the so-called stochastic background of GWs. The production of the primordial part of this stochastic background (relic GWs) is well known in literature starting from \cite{allen}, in which, using the so-called adiabatically-amplified zero-point fluctuation process, it has been shown in two different ways how the inflationary scenario for the early Universe can provide the signature for the spectrum of relic GWs. For the present model, such a signature is the strong suppression of $\rts$. However, other mechanisms for the generation of GWs, like cosmic strings \cite{Vi81,Vi811,Vi812,Vi813,Vi814} and string cosmology \cite{Ga93,Ga931,Ga932,Ga933,Ga934}, are possible and it becomes interest to analyze their propagation in the expanding Universe. In fact, the absence of the inflationary GW background favors the analyses of specific template tracking other generation mechanisms.

For a general derivation of the propagation equation of GWs in the extended framework, we start considering \refeq{nonan} without specifying, in the first instance, the form of $f(R)$. Using the conformal transformation $\tilde{g}_{\mu\nu}=e^{2\phi}g_{\mu\nu}$, with $e^{2\phi}=f\primo(R)$ ($\phi$ being is the conformal scalar field), and deriving the Einstein-like conformal equations, the GW evolution is described by $\tilde{\square}\,\tilde{h}_{i}^{j}=0$, in which $\tilde{\square}=e^{-2\phi}(\squareB+2\phi^{,\lambda}\nabla_\lambda)$. Since no scalar perturbation couples to the tensor part of GWs, one can easily recognize that the non-rescaled $h_{i}^{j}$ is a conformal invariant, since the following relation holds $\tilde{h}_{i}^{j}=\tilde{g}^{lj}\delta\tilde{g}_{il}=e^{-2\phi}g^{lj}e^{2\phi}\delta g_{il}=h_{i}^{j}$. As a consequence, the plane-wave amplitude $h_{i}^{j}=h(t)\;\epsilon_{i}^{j}\;e^{ik_{i}x^{i}}$ where $\epsilon_{i}^{j}$ is the polarization tensor, is the same in both metrics and this means that the background is changing while the tensor wave amplitude is not.

In order to study the cosmological stochastic background, the operator $\tilde{\square}$ can be specified for a RW flat model and the propagation equation becomes
\begin{equation}
\ddot{h}+[3H+2\dot{\phi}]\,\dot{h}+k^{2}a^{-2}\,h=0\;,\label{eq:10}
\end{equation}
being $\squareB=\partial^{2}_t+3H\partial_t$ and $k=k_i k^i$ the comoving wave number. Using a power-law behavior for the scale factor ($a=\azero\;[t/\tzero]^x$ with $x<1$) and the present $f(R)$, we get
\begin{align}
\phi(t)%&=\frac{1}{2}\ln\left[1+nqR^{n-1}\right]=\nonumber\\
=\tfrac{1}{2}\ln\big[1+nq\big(6x(2x-1)\big)^{n-1} \,t^{-2(n-1)}\big]\;.
\end{align}
During the radiation-dominated era ($x=1/2$), $\phi(t)$ vanishes identically. Thus, \refeq{eq:10} reduces to its GR form and the standard result for the GW propagation follows, \emph{i.e.} $h\propto \sin({k \eta})/k\eta$, where $\eta$ is the conformal time, defined by $dt = ad\eta$. For $x\ne1/2$, $\dot\phi$ is given by
\begin{align}
\dot\phi%&=\frac{n(1-n)q\big[6x(2x-1)\big]^{n-1} \,t^{1-2n}}{1+nq\big[6x(2x-1)\big]^{n-1} \,t^{-2(n-1)}}=\nonumber\\
%=\frac{(n-1)}{t}\frac{1}{(t/t_*)^{2(n-1)}-1}\;,
=[(n-1)/t]\;[(t/t_*)^{2(n-1)}-1]^{-1}\;,
\end{align}
where we have defined $t_*=(n|q|)^{{1}/{2(n-1)}}\left|6x(2x-1)\right|^{1/2}$. It is easy to see that for $t\ll t_*$, $\dot{\phi}\simeq(1-n)/t$. Since $H\simeq 1/t$, in this regime $\dot\phi$ and $H$ are comparable and $f(R)$ can affect the propagation of GWs. When $t\sim t_*$, $\dot\phi$ becomes very large (and eventually diverges for $t=t_*$) and the corrections to the EH action dominate the dynamics. Finally, when $t\gg t_*$, we have that $|\dot\phi|\ll H$ and the standard behavior is recovered.

In the regime $t\ll t_*$, \refeq{eq:10} can be recast, using time variable $u\equiv k\eta$, as
\begin{equation}
\frac{d^2h}{du^2}+\frac{2(x-n+1)}{(1-x)u}\frac{dh}{du}+ h=0\;.
\label{eq:13a}
\end{equation}
Since $\eta$ represents the cosmological horizon, $u=1$ roughly corresponds to the time of horizon crossing and $u\ll 1$ ($u\gg1$) represents a wave that is far outside (inside) the horizon. The general solution to this equation reads
\begin{align}
\label{eq:14}
h(u)=u^\alpha \left[A_{1}J_{\alpha}(u)+A_{2}Y_{\alpha}(u)\right]\;,
\end{align}
where $\alpha=(2n-3x-1)/(2(1-x))$, $J_{\alpha}$ and $Y_\alpha$ are the Bessel functions of the first and second kind, respectively, and $A_{1,2}$ are integration constants. In the range of interest ($2<n<3$ and $0<x<1$), we have that $\alpha > 3/2$ and we can safely use the expansions for the Bessel functions. The two independent solutions in \refeq{eq:14} are both well-behaved when $u\to0$. In fact, $u^\alpha J_\alpha(u)\to 0$, while $u^\alpha Y_\alpha(u)$ tends to a constant value. On the other hand, when $u\gg 1$ both solutions oscillate with an amplitude that grows with time like $u^{\alpha-1/2}$. This is in striking contrast with the standard GR behavior, where the amplitude decays with time as $h\propto \sin(u)/u$.

These results make relevant to estimate when $t_*$ takes place in the cosmological history. The value of $L_q$ is bounded from above by the requirement of a matter-dominated era. The exact value of $L_q^{Max}$ depends only weakly on $n$ so that we cna consider $L_q\lesssims70\pc$ and we find that $t_* \lesssims400\,\mathrm{yr}$; since the matter-dominated era began at $t\simeq 10^{5}\,\mathrm{yr}$, we can conclude that during this epoch the propagation of GWs results to be standard. Finally, it can be easily shown that the same behavior occurs also in a de Sitter phase.

\section{The weak-field limit and the massive boson issue}
We now want deepen a subtle question, concerning the real nature of the scalar degree of freedom present in our model, especially in view of the comparison with the analytical case. If we assign an analytical form $f(R)=R+g(R)$ for the model, in the weak-field limit, \emph{i.e.} when the spacetime curvature is very small ($R\sim0$), we can take the following Taylor approximation: $g(R)\simeq g\primoprimo(0)R^2+...$. Thus, setting $g\primoprimo(0)\equiv6/\mu^2$, the associated scalar-tensor theory contains a massive boson field of mass $\mu$, see \cite{DK03,SH07,C03,CDTT04}. When solving the stationary and non-stationary Einstein equations, near the Minkowski spacetime, this massive boson emerges also in the Jordan frame as a Yukawa correction to the Newtonian term and a massive non-traceless mode of the gravitational wave, respectively. In particular, the trace part of the gravitational field can be summarized by the following equation for the weak-field scalar of curvature
\begin{equation}
\Box R = \mu^2R\;.
\label{dlr}
\end{equation}

In the present model, the function $g(R)=qR^n$ is not analytical and therefore it is not expandable around $R\sim H^2_0\sim0$. Apparently, we should have
\begin{equation}
\Big(\frac{\mu}{m_\mathrm{pl}}\Big)^{2}\simeq \frac{6\;\textrm{sign}[q]}{n(n-1)}\;
\Big(\frac{L_H}{L_q}\Big)^{2n-2}\mathcal{O}(10^{-120})\;, 
\label{tacm}
\end{equation}
where $L_H\simeq H_0^{-1}\sim 10^{27}$cm denotes the present Hubble scale. Recalling the range of variability for the lenght-scale $L_q$ and for $n\to3$, this square ratio is at most $\mathcal{O}(10^{-60})$. Furthermore, since we have to require that the parameter $q$ takes negative values and $n=(2\ell+1)/(2m+1)$ in order to deal with a de Sitter phase of the Universe, we would have the very unpleasant feature of a tachyonic mass ($\mu^2<0$) which would invalidate the considered $f(R)$ model. However, the situation is rather different in view of the non-analytical character of the adopted $g(R)$. In fact, while the second derivative is still defined in $R\sim0$, this is not true for all the subsequent higher order derivatives. In this sense, the huge and negative value of the parameter $\mu^2$ is here meaningless from a physical point of view. Our theory can not predict a (linear) free massive boson field, simply because it is intrinsically non-linear also in the weak-field limit.

Our modified perturbation equations of the Minkowski space contains, in addition to the ordinary linear GR terms, small non-linear contributions which are however greater than the quadratic terms in the perturbation amplitude (neglected in standard linear GR). This feature is a consequence of the non-analytic nature of the present model and of the restriction $2<n<3$. The obtained weak-field amplitudes, both in the static spherically symmetric case \cite{LM09} and in the non-stationary ripple propagation \cite{FLM11}, increase with the radial and the temporal variables, and the validity if the obtained profile is restricted to well-defined domains determined by the free parameters of the model. By other words, our weak-field limit is non-linear but perturbative and, when it stops to be valid, a real non-perturbative solution must be evaluated which however can have all the regularity requirements to describe a physical spacetime.

Nonetheless, from a pure phenomenological point of view, our scalar mode is a (peculiar) massive boson, since, as shown in \cite{FLM11}, the velocity of the non-linear non-traceless mode propagates with a speed less than that of the light. The associated propagation equation for the scalar of curvature takes the form
\begin{equation}
\Box R^{n-2} = \frac{1}{3nq}\;R\;,
\label{peq}
\end{equation}
which, in the limit $n\to2$, approaches the correct linear analytical case although this value can not be consistently included in the present model (the solution of the non-analytical ($2<n<3$) and analytical ($n=2$ and $n=3$) cases cannot be continuously matched together, belonging to distinct sectors of modified gravity, see \cite{LM09}).

We conclude observing that, despite the features of the non-analytical Lagrangian we considered here, the scalar-tensor representation offers a corresponding peculiar but well-defined picture in terms of the scalar field stability. The constraints $q<0$ and $n=(2\ell+1)/(2m+1)$, coming from the viability of the de Sitter phase, implies that, for the reality of the potential, the scalar field is restricted to take values in the positive real axis, see \figref{FIGV}. The potential profile is then a monotonic increasing function from $\varphi=0$, which results to be a stable configuration though the mass is there not well defined (the second derivative of $V_+(\varphi)$ diverges in $\varphi=0$). This fact reflect the non-linear nature of the massive boson near its minimum, but ensures that GR is contained in the theory as the proper limit $R\to0$.

\section{Concluding remarks}\label{sette}

We have considered a modified theory of gravity of the form $f(R)=R+qR^n$, where both $q<0$ and $2<n<3$ are free parameters. We have assumed a power-law behavior for the cosmological scale factor which allows to represents both the radiation- and matter-dominated eras. As a result, an upper bound of $\sim 70\pc$ for the characteristic length scale of the model was found by guaranteeing the existence of this epoch. The presence of this cosmological bound and of the Solar-System constraints, has allowed us to define a restricted region of the parameter space. Moreover, we have shown how an exponential early evolution of the Universe is still present in the extended $f(R)$ scheme. The expansion rate was found to be much smaller than the standard value by several orders of magnitude due to the presence of a minimum length scale. Thus we get an inflationary scale arbitrarily close to the Planck one without violating the present limits on the tensor-to-scalar ratio, since the latter is proportional to the expansion rate and gets correspondingly small values. The observational implication is that, if the gravitational action is of the form considered here, tensor modes will not be detected even by next-generation CMB-polarization experiments. On the other hand, a detection of tensor modes in the near future would falsify the present model. It worth noting that, for $n>3$ the presented scheme would be significantly altered. This is a consequence of the modification suffered by the weak field limit, more than of an intrinsic change of the cosmological setting which remains isomorphic to the one discussed in Sections \ref{tre} and \ref{quattro}. In fact, for such a case, the corrections coming from the modified Lagrangian are, in the weak field limit, smaller than the quadratic correction that GR itself contains. As a result, the low bounding value for $L_q$ can no longer be estimated as in Eq.\reff{LMin} and it can be expected to be seriously lowered. In this sense, for $n>3$ it is no longer possible to claim the viability of the treated $f(R)$ model, along the paradigm here developed. The issue of the GW propagation on a RW background has also been addressed, showing that no significant modifications come out with respect to the standard case. Therefore, if the tensor-to-scalar ratio is strongly suppressed \emph{ab initio} there is no chance that the later propagation of GW amplifies its value.

\vspace{0.5cm}
{\footnotesize This work has been partially developed in the framework of the \emph{CGW collaboration} (www.cgwcollaboration.it). The work of M. L. has been supported by the PRIN grant \emph{Galactic and extragalactic polarized microwave emission} (Contract No. PRIN 2009XZ54H2-002).}

\end{document}